# Multivariate Techniques for Identifying Diffractive Interactions at the LHC

Mikael Kuusela[1,2], Jerry W. Lämsä[1,3,4], Eric Malmi[1,2], Petteri Mehtälä[1,3], and Risto Orava[1,3,5]

Wednesday, September 16, 2009

Abstract

Close to one half of the LHC events are expected to be due to elastic or inelastic diffractive scattering. Still, predictions based on extrapolations of experimental data at lower energies differ by large factors in estimating the relative rate of diffractive event categories at the LHC energies. By identifying diffractive events, detailed studies on proton structure can be carried out.

The combined forward physics objects: rapidity gaps, forward multiplicity and transverse energy flows can be used to efficiently classify proton-proton collisions. Data samples recorded by the forward detectors, with a simple extension, will allow first estimates of the single diffractive (SD), double diffractive (DD), central diffractive (CD), and non-diffractive (ND) cross sections. The approach, which uses the measurement of inelastic activity in forward and central detector systems, is complementary to the detection and measurement of leading beam-like protons.

In this investigation, three different multivariate analysis approaches are assessed in classifying forward physics processes at the LHC. It is shown that with gene expression programming, neural networks and support vector machines, diffraction can be efficiently identified within a large sample of simulated proton-proton scattering events. The event characteristics are visualized by using the self-organizing map algorithm.

[1] *Helsinki Insitute of Physics, PL 64 (Gustaf Hällströmin katu 2a), FI-00014 University of Helsinki, Finland*
[2] *Adaptive Informatics Research Centre, Helsinki University of Technology, FI-02015 TKK, Espoo, Finland.*
[3] *Division of Elementary Particle Physics, Department of Physics, PL 64 (Gustaf Hällströmin katu 2a), FI-00014 University of Helsinki, Finland*
[4] *Iowa State University, Physics Department, Ames, Iowa 5001, U.S.*
[5] *Presently at CERN, CH-1211 Geneva 23, Switzerland*

# 1 Road map to forward physics studies at the LHC

Proton-proton event categories

High energy proton-proton collisions are here divided into five different categories[6]: elastic (EL), single diffractive (SD), double diffractive (DD), central diffractive (CD) and non-diffractive scattering (ND) (see Figure 1). The sum of these categories defines the total pp cross section as:

$$\sigma_{TOT} \equiv \sigma_{EL} + \sigma_{SD} + \sigma_{DD} + \sigma_{CD} + \sigma_{ND}. \qquad (1)$$

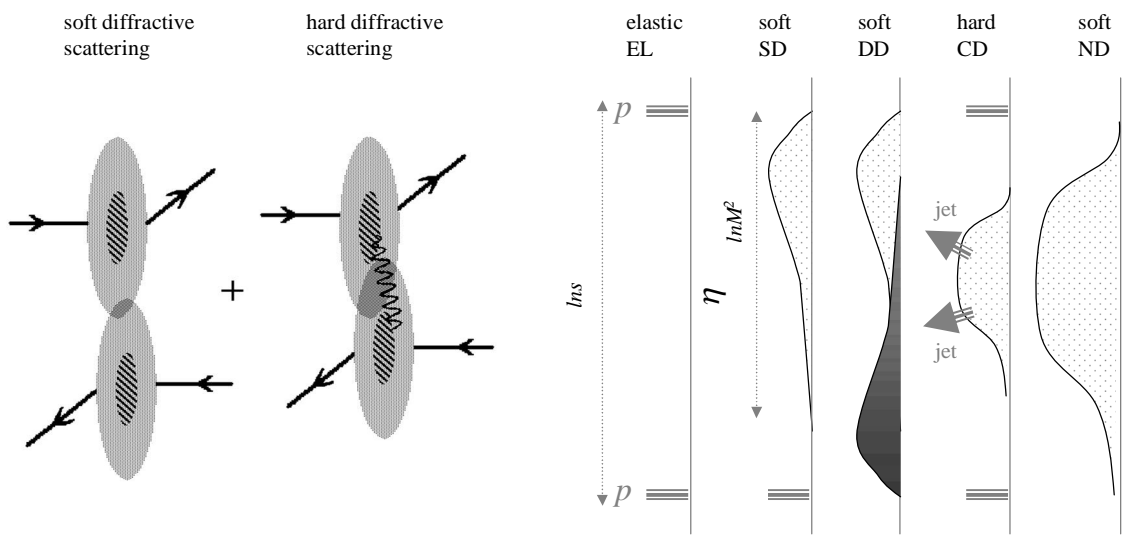

**Figure 1.** *Classification of proton-proton collisions into elastic (EL), single diffractive (SD), double diffractive (DD), central diffractive (CD) and non-diffractive (ND) processes. The diffractive events (soft and hard) are viewed in the s-channel with their characteristic pseudorapidity distributions shown on the right hand side. Hard diffractive scattering exhibits a hard scale shown here as jet production within the hadronic systems. Diffractive scattering can be viewed as giving a "snap-shot" of the instantaneous (transverse) parton configuration of the (Lorentz-contracted) proton during the interaction[7]*

Inelastic diffractive scattering can be understood by viewing an interacting proton as a superposition of different states that undergo unequal absorption [1]. These "transmission eigenstates" are sometimes identified as configurations of parton constituents inside a proton [2,3]. In single diffractive scattering, either proton-1 (circulating anti-clockwise around the LHC ring) *or* proton-2 (circulating clockwise around the LHC ring) gets excited into a diffractive state with mass $M$ ($M \geq m_p + m_\pi$)[8]. A single diffractive event is characterized by a large ($\Delta\eta > 2$) rapidity gap that in diffractive excitation of proton-1 (SD1) separates the diffractively produced cluster of particles from proton-2 which remains intact. In a symmetric case, proton-2 is diffractively excited (SD2), and the rapidity gap spanning between proton-1 and the diffractive system identifies the event. A description of beam-1 (beam-2) diffraction is given by the double differential cross section $d^2\sigma_{SD1}/dM_1^2 dt$ ($d^2\sigma_{SD2}/dM_2^2 dt$), which can be considered as a probability of the beam-1 proton (beam-2 proton) to get excited into a diffractive state of mass $M_{1,2}$ with a four-momentum transfer squared of $-t_{1,2} = -(p_{1,2} - p_{M1,2})^2$, where $p_{1,2}$ denote the four-momenta of the beam-1,2 protons and $p_{M1,2}$ the four-momenta of the diffractively excited beam-1,2 systems. The double diffractive cross section is given as $d^3\sigma_{DD}/dM_1^2 dM_2^2 dt$. As in all diffractive processes, the t-dependence is usually given as an exponential, $d\sigma/dt \propto \exp(-Bt)$.[9] The diffractive mass grows rapidly after the threshold ($M = m_p + m_\pi$) is reached, oscillates in the N* resonance region, and subsequently falls as $d\sigma/dM^2 \propto 1/M^2$. The central diffractive cross section is given as $d^3\sigma_{CD}/dM^2 dt_1 dt_2$, where $M$ is the invariant mass of the centrally produced diffractive system.

---

[6] Beam related backgrounds due to beam-gas and beam halo induced interactions are removed by using their specific signatures, see Ref. [15].

[7] The interaction time, $\tau_{in}$, between a pair of colliding protons can be estimated as $\tau_{in} \approx 2R_p / \gamma_{cm} \cong 4R_p m_p / \sqrt{s}$, where $R_p$ is the proton dimension (~ 0.5 fm). The Lorentz contraction flattens the proton as $1/\sqrt{s}$. Since $R_p/c \ll \tau_{in}$, the proton appears to be 'frozen' into its instantaneous (transverse) parton configuration.

[8] Diffractive processes leading to small ($M < 10\ GeV$) or large ($M \geq 10\ GeV$) diffractive masses are a subject of a separate study, see Ref. [5].

Access to different classes of proton-proton scattering events at the LHC is severely impaired by the limited forward acceptance of the base line experiments. Usually only about one half of the total pp cross section, $\sigma_{TOT}$, is expected to be seen during the nominal LHC running conditions [4]. This limits the accuracy with which any absolute cross section can be determined. Together with simple extensions of the planned forward detector systems [5-7], a technique for efficient and pure classification of the recorded data would allow a major improvement in physics analysis at the LHC.

In case of hard diffractive scattering[10], the forward detector coverage directly relates to the accessible region of parton fractional momenta. The fractional momenta of initial state partons $x_{1,2}$ are connected to the pseudorapidities $\eta_{1,2}$ and transverse energies $E_T$ of the observed jets through the relation:

$$x_{1,2} = \frac{E_T}{\sqrt{s}} \exp\left[\pm \frac{\eta_1 + \eta_2}{2}\right]. \quad (2)$$

In diffractive proton-proton scattering models inspired by Good and Walker (see Refs. [2,3]), the expected number of hard interactions in a proton-proton collision with impact parameter b is determined by the instantaneous parton configurations within the beam-1 and beam-2 protons. If the configuration is dominated by a few hard partons, such as the valence quarks, the likelihood of hard scattering is suppressed. In case these parton configurations are dominated by soft gluons with their fractional energies of the order $x \sim M_{min}^2/s$, the number of hard interactions is enhanced. It is, therefore, of high priority to extend the forward rapidity coverage as much as possible.[11]

---

[9] $\Delta\sigma/dt \sim \exp(-Bt)$ with slope values between $B = 6 - 10~GeV^2$ are used in different models.
[10] The diffractive processes with a hard scale, i.e. with jets, heavy quarks, W/Z or Higgs bosons typically have rates of the order of one percent as compared to the corresponding inclusive high $E_T$ pp interactions [8].
[11] The acceptance in $x$ is based on acceptance of the diffractive mass: $x \geq (M_{min}/\sqrt{s})^2 \approx (1~GeV/14~000~GeV)^2 \approx 10^{-8}$.

Constraints

A set of general constraints connect the total and diffractive proton-proton cross sections together. Using these - and an independent luminosity measurement - together with an accurate event classification method, the diffractive cross sections can be determined.

The sum of elastic and diffractive cross sections is expected, in certain approxiamation, to satisfy the Pumplin bound [9] (Fig. 2):

$$\sigma_{EL} + \sigma_{SD} + \sigma_{DD} \leq \tfrac{1}{2}\, \sigma_{TOT} \qquad (3)$$

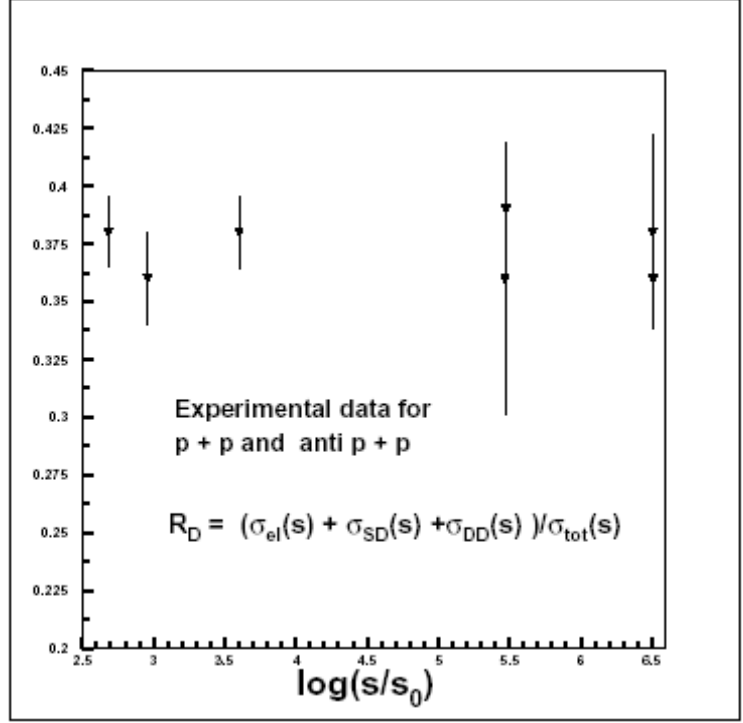

**Figure 2** *Experimental data for $R_D=(\sigma_{EL}+\sigma_{SD}+\sigma_{DD})/\sigma_{TOT}$ in $pp/\bar{p}p$ [10].*

and (neglecting absorption effects) obey the factorization rules:

$$\frac{d^4\sigma_{CD}}{dt_1 d\xi_1 dt_2 d\xi_2} = \frac{d^2\sigma_{SD}}{dt_1 d\xi_1} \frac{d^2\sigma_{SD}}{dt_2 d\xi_2} \frac{1}{\sigma_{TOT}} \qquad (4)$$

$$\frac{d^4\sigma_{DD}}{dt_1 dM_1^2 dt_2 dM_2^2} = \left(\frac{d^2\sigma_{SD}}{dt_1 dM_1^2} \frac{d^2\sigma_{SD}}{dt_2 dM_2^2}\right) \Big/ \left(\frac{d\sigma_{EL}}{dt}\right), \qquad (5)$$

where $t_{1,2}$ is the 4-momentum transfer squared: $-t_{1,2} = -(p_{1,2} - p_{1,2}')^2$, $\xi_{1,2} = 1 - |\vec{p}\,'_{1,2}|/|\vec{p}_{1,2}|$ denote the momentum loss fractions of the two scattered protons, and $M_{1,2}$ the invariant mass of the diffractive system.

The cross section for dissociation of a beam-1,2 proton into a diffractive system with mass $M$ has the approximate form [11]

$$\sigma_{SD1,2} = \int \left( M^2 \frac{d\sigma_{SD1,2}}{dM^2} \right) \frac{dM^2}{M^2} \approx \lambda(\ln s)\sigma_{EL}, \quad (6)$$

where $\lambda \equiv g_{3P}/g_N$, with $g_{3P}$ the triple-Pomeron coupling and $g_N$ the coupling of the Pomeron to the proton. The $ln(s)$ 'rapidity' factor comes from the integration $\int dM^2 / M^2$.

In Figure 3, the experimental results on the ratio $\sigma_{EL}/\sigma_{TOT}$ are shown as a function of the centre-of-mass energy.

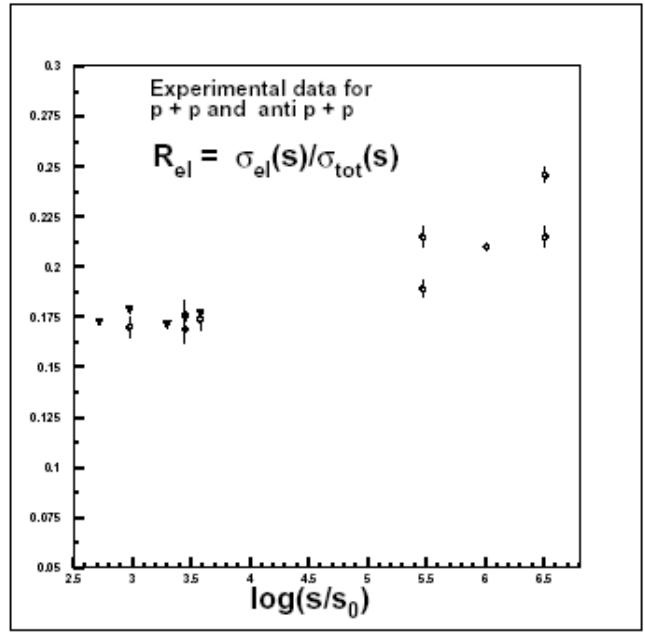

**Figure 3** *Experimental data for $\sigma_{el}/\sigma_{tot}$ in pp/ $\bar{p}$ p [10].*

An extrapolation of the ratio $\sigma_{EL}/\sigma_{TOT}$ (Fig. 3) to the LHC energies gives $\sigma_{EL} \approx 0.22 \cdot \sigma_{TOT}$. Assuming a total pp cross section of *90 mb* [12], $\sigma_{EL} \approx 20$ mb. Eq. 6 with $\lambda \approx 0.023$ gives for the single diffraction $\sigma_{SD1,2} \approx 5$ mb. The elastic and inelastic diffraction represent about *35-40 %* of the total pp cross section (Fig. 2), and the following "Good and Walker inspired" cross section estimates are obtained for the LHC energy of $\sqrt{s} = 14$ TeV: $\sigma_{EL} \approx 20$ mb, $\sigma_{SD1} + \sigma_{SD2} \approx 10$ mb, and $\sigma_{DD} \approx 5$ mb. Finally, using factorization (Eq. 4), an estimate for large mass central diffraction is obtained as:

$$M^2 \frac{d\sigma_{CD}}{dM^2 d\eta} \approx \frac{1}{\sigma_{INEL}} \left( \frac{d\sigma_{SD}}{d\Delta\eta_{gap}} \right)^2 \approx 1-10\mu b, \quad (7)$$

where $\eta$ is the central rapidity location of the diffractively produced system and $\Delta\eta_{gap}$ the extent of rapidity gap used to define the events as diffractive.

A summary of recent theoretical calculations [13,14], together with the PYTHIA6.205, PHOJET1.12 simulation results (see Ch. 4) and the simple "Good and Walker inspired" estimate (see above) for the diffractive pp cross sections are given in Table 1.

**Table 1.** P*redictions for the proton-proton cross sections at the LHC ($\sqrt{s}$ = 14 TeV), GLMM and KMR from Refs. [13,14], for PYTHIA6.205 and PHOJET1.12 see Ch. 4, for "GW" see the text.*

| Cross Section (mb) | GLMM (mb) | KMR (mb) | PYTHIA6.205 | PHOJET1.12 | "GW"(mb) |
|---|---|---|---|---|---|
| $\sigma_{TOT}$ | 92.10 | 88.00 | 101.50 | 119.00 | 90.00 |
| $\sigma_{EL}$ | 20.90 | 20.10 | 22.20 | 34.40 | 20.00 |
| $\sigma_{SD}$ | 11.80 | 13.30 | 14.30 | 11.00 | 10.00 |
| $\sigma_{DD}$ | 6.10 | 13.40 | 9.80 | 4.06 | 5.00 |
| $(\sigma_{EL} + \sigma_{DIFF})/\sigma_{TOT}$ | 0.42 | 0.53 | 0.46 | 0.42 | 0.39 |

## 2  Detectors relevant to forward physics

Forward detectors of the LHC experiments are designed to maximally cover forward physics processes with varying beam conditions. While the aim of the TOTEM Collaboration is to measure proton-proton elastic scattering, total cross section and soft diffraction by using dedicated beam optics conditions with relatively large, unsqueezed (large $\beta^*$) beams [15], the primary goal of the ATLAS[12] and CMS[13] is to observe new physics phenomena with hard energy scales requiring high luminosities and tightly focused (low $\beta^*$) beams with a non-zero crossing angle. The ALICE and LHCb[14] experiments will complement the forward physics studies at lower diffractive masses.

A special TOTEM high-$\beta^*$ beam optics ($\beta^*$ = 90 m) was found to have a good acceptance of elastically scattered protons (-$t_{min}$ ≥ $10^{-2}$ $GeV^2$) and a coverage of basically *all* the diffractive protons[15] ($\xi \geq 10^{-8}$). These optics conditions are hoped to be available during the set-up period of the LHC beams, since no special injection optics settings are required as is the case with the nominal TOTEM high-$\beta^*$ optics ($\beta^* \approx 1500$ m). The leading diffractive protons are planned to be measured by a set of Roman Pot stations at ±147m and ±220m from IP5 (Fig. 4.1), and the forward particle flows by small angle spectrometers, T1 and T2, placed at *3.1 ≤|η| ≤ 4.7* and *5.2 ≤|η| ≤6.5* symmetrically on both sides of IP5 (Fig. 4.2).

---

[12] For the ATLAS forward physics plans see [16]; for an early study of the potential of forward physics in conjunction with the ATLAS experiment, see [7].
[13] For a study of the Forward Shower Counters in facilitating forward physics at the CMS experiment, see [5]
[14] For a study on central diffraction at the LHCb experiment, see [6].
[15] The acceptance in $\xi$ is based on acceptance of the diffractive mass: $\xi \geq (M/\sqrt{s})^2 \approx (1\ GeV/14\ 000\ GeV)^2 \approx 10^{-8}$.

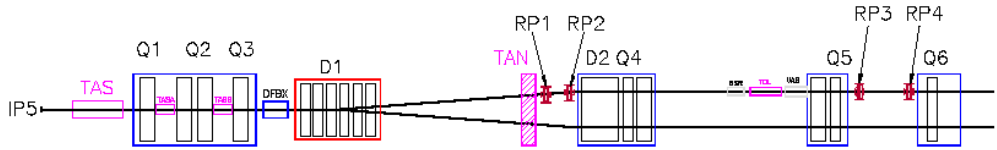

**Figure 4.1** The *TOTEM lay-out of leading proton detectors (Roman Pots). The detector locations at ±147m (RP1) and at ±220 m (RP3) are shown [15]*.

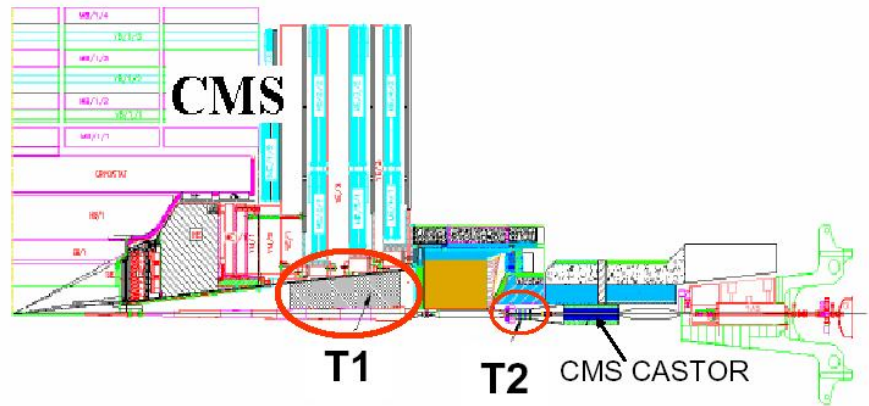

**Figure 4.2** *The lay-out of TOTEM inelastic spectrometers, T1 based on Cathode Strip Counters (CSC), and T2 based on Gas Electron Multipliers (GEMs) [15]*.

Scintillation counters and gaseous electron multipliers (GEMs) could be added to surround the beam pipes, with *60m < |z| < 85m*, and at further locations out to *±140m* on both sides of the LHC interaction point (IP5), see Fig. 4.3 [5]. These would detect showers from very forward particles interacting in the beam pipe and surrounding material. These detectors can be used to make measurements of rapidity gaps in the absence of significant pile-up, i.e. multiple events produced in a given bunch crossing. Such detectors could be deployed as modest upgrades to the existing base line experiments at the LHC [5-7]. During the early phase of LHC operation, the bunch intensities will be low and practically no pile-up events are expected, so that a sample with single interaction events can be isolated. Typically when the average number of events per bunch crossing is $\langle N_{ev} \rangle \approx 1$ (for example for $6 \cdot 10^{10}$ particles per bunch, with *N = 156* bunches in the machine and *β\** of *~2m*) about *40%* of the bunch crossings will have exactly one event. Multiple events in bunch crossings can be distinguished due to the excellent tracking capabilities of CMS, where multiple primary vertices can be measured.

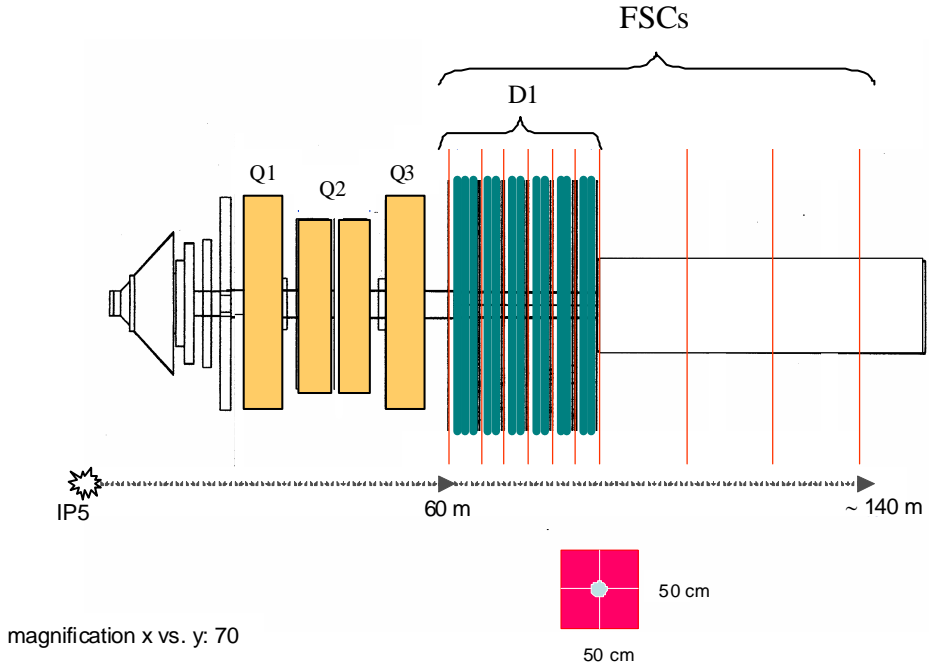

**Figure 4.3** *The proposed upgrade of Forward Shower Counters, FSCs, to the CMS forward detector lay-out at ±60 to ±140 meters from the IP, the 10 vertical lines from ±60 m on indicate the locations of the proposed veto counters [5].*

The ATLAS ALFA detector focuses on the absolute luminosity calibration for the different relative luminosity monitors of the experiment and specifically for the LUCID monitor [16]. The detector is designed to run only for a short period and under special beam conditions at low luminosity. In order to reach the Coulomb interference region at $t \approx 6.5 \cdot 10^{-4}\ GeV^2$, LHC run conditions based on the beam optics with very high $\beta^*$, $\beta^* \approx 2600\ m$ at luminosity, $L \approx 10^{27}\ cm^{-2}s^{-1}$ and low emittance are used.

The ALFA detector is located at *±240m* on both sides of the ATLAS interaction point. In each location, two vertical Roman Pot stations are installed at a distance of *4m* from each other.

A forward physics scenario that can be used in the case diffractive protons are *not* measured, is based on recording inelastic activity by the forward detectors. In conjunction with the central ATLAS, ALICE, CMS, or LHCb detectors, powerful event selection algorithms can be constructed for measuring diffractive cross sections. It is expected, that it is with these measurements that the first experimental estimations of diffractive cross sections will become available. The multivariate methods of this paper are studied under the assumption that the diffractive protons are not detected and the proposed event selection methods will complement the eventual leading proton measurements planned to be carried out by TOTEM, CMS and ATLAS experiments.

The naïve extrapolation discussed above (Table 1, p.6) gives for the nominal LHC design luminosity of $10^{34}\ cm^{-2}s^{-1}$, a mean number of $N_{TOT} = 30$ proton-proton collisions, with $N_{EL} = 7$ elastic, $N_{SD} = 3$ single diffractive (*1.5* SD1 and *1.5* SD2 events), $N_{DD} = 2$ double diffractive, and $N_{ND} = 18$ inelastic non-diffractive (ND) events with each beam crossing in every *25 ns*. The cross section for central diffractive events (CD) is highly uncertain, and could be anywhere between *1 µb* to *1 mb* [13,14,19]. The probability to observe a single LHC event of a given event category, i.e. without contamination of others of the same type, is given by *exp(-$N_i$)*. In case of the initial measurements, the run conditions are relaxed, with practically no pile-up events until luminosities of *$10^{31}$ $cm^{-2}s^{-1}$*. Since about *50%* of the total pp cross section is likely to be built up by diffractive processes, large samples of events will be collected in a few days of running. Simultaneously, non-diffractive processes and machine induced backgrounds (beam-gas and beam-halo) could be studied in detail.

# 3  Characteristics of pp event classes

Characteristics of the different event categories are best seen in the rapidity space (Figures 5), where distribution of the final state particles and their transverse energies in pseudorapidity reveal the distinct dynamics of each type of a scattering process.

## Single diffractive events

In single diffraction, either the beam-1 proton (circulating anti-clockwise in the LHC ring) or the beam-2 proton (circulating clockwise in the LHC ring) is excited to a higher mass state with the proton quantum numbers. Relatively few particles carry the total collision energy at small polar angles as seen in Figure 5.1, in which the beam-1 proton undergoes diffraction scattering and ends up as an excited high mass state on the right hand side of Fig. 5.1. The main energy flows are seen by the CMS forward calorimeters, Hadron Forward (HF), CASTOR and Zero Degree Calorimeter (ZDC)[16]; the multiplicity flows are seen in the TOTEM T1 and T2 detectors and the proposed Forward Shower Counters (FSCs).

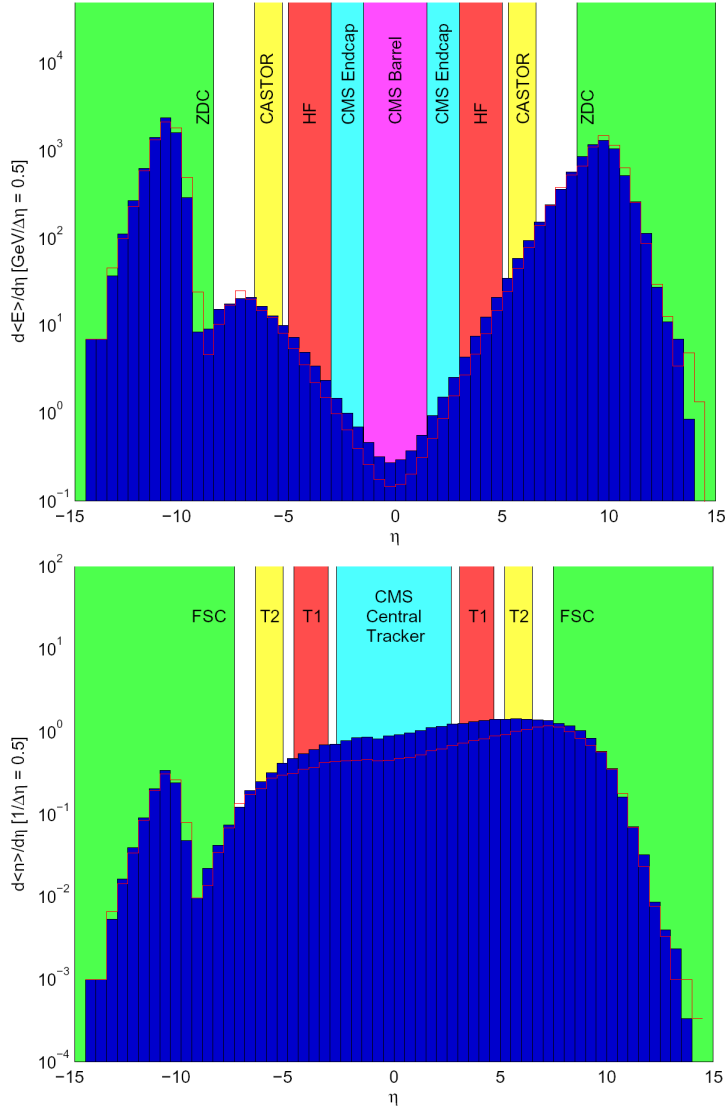

**Figure 5.1** *The energy flows (upper figure) and multiplicities (lower figure) as predicted by the PYTHIA (histograms) and PHOJET (broken lines) Monte Carlo models in an average single diffractive event with the beam-1 proton (seen in the right hand side of the figures) excited into a higher mass state with mass M (SD1) at the LHC energy (14 TeV).The separate (coloured) regions indicate the acceptances of the CMS/TOTEM detectors relevant for the energy (upper figure) and multiplicity (lower figure) measurements. Note the logarithmic vertical scale.*

---

[16] Note that the Zero Degree Calorimeters are placed at *0* degrees at the point where the LHC beams separate, i.e. they only detect hard neutral particles, such as neutrons, gammas, etc.

## Double diffractive events

In double diffractive events, both the beam-1 proton and beam-2 proton are excited into higher mass states and are seen to populate both sides of the rapidity space in Fig. 5.2. The two diffractive systems release their main energy flows at rapidities in excess of $|\eta| > 5$, with a number of soft particles filling up the central rapidities.

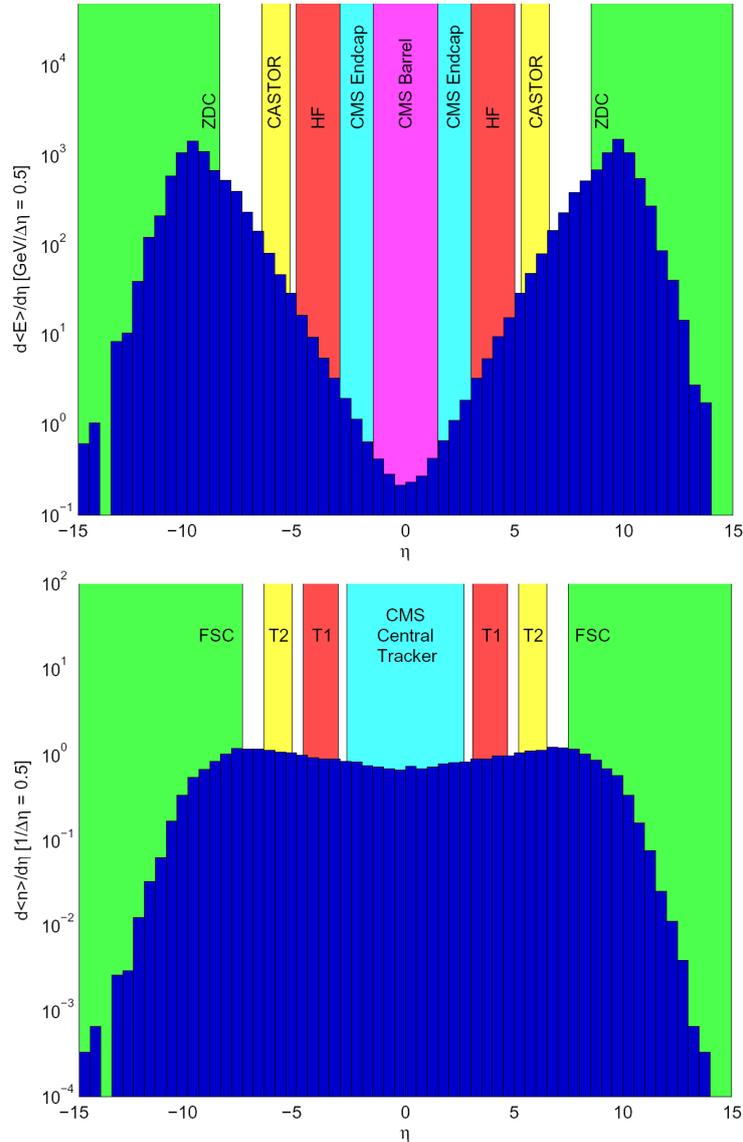

**Figure 5.2** *The predicted (PHOJET) energy flows (upper figure) and multiplicities (lower figure) in an average double diffractive (DD) event at the LHC energy (14 TeV). Both beam-1 proton (ending up on the right hand side of the figure) and the beam-2 proton (ending up on the left hand side of the figure) are excited to diffractive states with masses $M_1$ and $M_2$.. The separate (coloured) regions indicate the acceptances of the CMS/TOTEM detectors relevant for the energy (upper figure) and multiplicity (lower figure) measurements. Note the logarithmic vertical scale.*

## Central diffractive events

In CD events, a diffractive system with central pseudorapidity and vacuum quantum numbers, predominantly with spin-parity $J^{PC} = 0^{++}$, is created in an exchange process between the beam-1 and beam-2 protons.[17] The central diffractive events are characterized by the relatively soft central, $|\eta| < 7$, and hard forward, $|\eta| > 8$, energy flows registered by the central and forward calorimeters. It is important to note the key role that the ZDC plays – in case of fast beam-like neutrals - in discriminating the central diffractive events. In case of the multiplicity flows, it is the proposed FSC system that gives the most crucial tagging information.

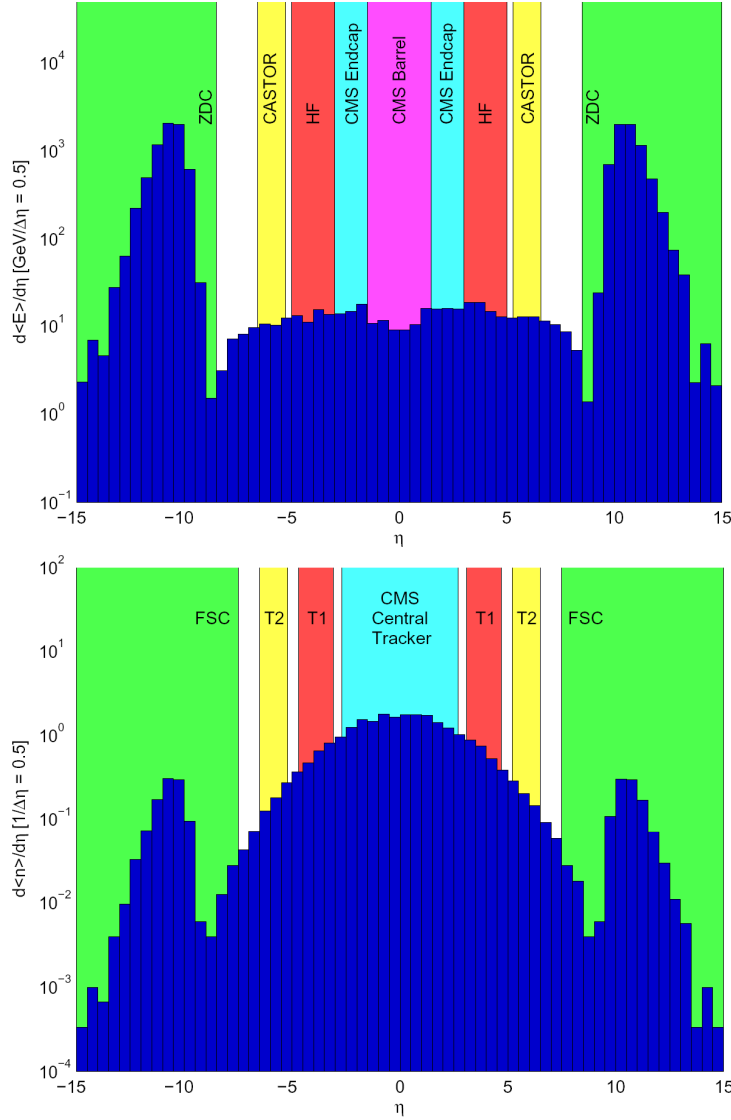

**Figure 5.3** *The predicted (PHOJET) energy flows (upper figure) and multiplicities (lower figure) in an average central diffractive (CD) event at the LHC energy (14 TeV). A central diffractive state with mass M and vacuum quantum numbers, predominantly spin-parity $J^{PC} = 0^{++}$, is created. Both beam-1 and beam-2 proton could get excited into a higher mass state as in single or double diffraction. The separate (coloured) regions indicate the acceptances of the CMS/TOTEM detectors relevant for the energy (upper figure) and multiplicity (lower figure) measurements. Note the logarithmic vertical scale.*

---

[17] There is a finite probability for either beam-1 or beam-2 proton, or both, to be diffractively excited as in the case of single or double diffractive processes.

## Non-diffractive events

In non-diffractive scattering, colour exchange between the beam-1 and beam-2 protons induces multiple particle production at central rapidities. As a result, the kinematically allowed phase space region is filled up by particles and radiation quanta as shown by PYTHIA simulated ND events in Figure 5.4. The CMS forward calorimeters, HF, CASTOR and ZDC cover the main part of non-diffractive energy flows carried by the relatively few forward going particles. Main multiplicity flows are registered by the central CMS tracking detectors, TOTEM T1 and T2 spectrometers and the proposed FSC counters.

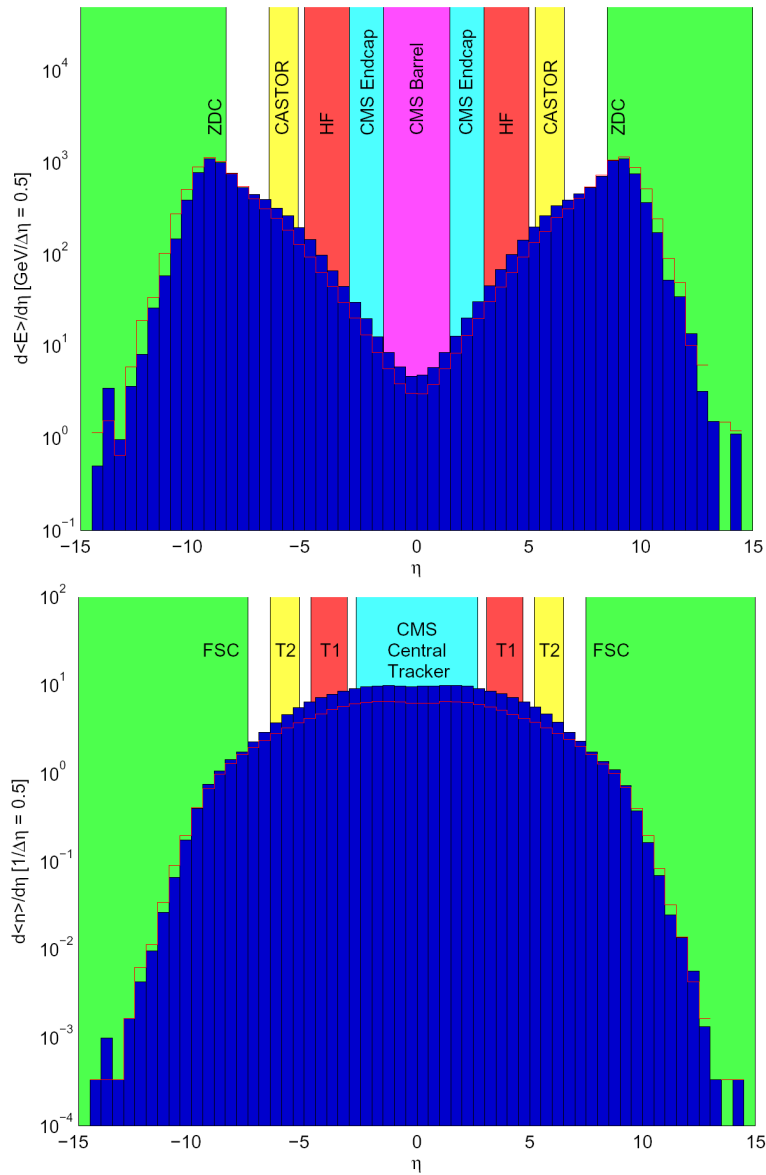

**Figure 5.4** *The energy flows (upper figure) and multiplicities (lower figure) predicted by the PYTHIA (histograms) and PHOJET (broken lines) Monte Carlo Models in an average non-diffractive (ND) event at the LHC energy (14 TeV). The separate (coloured) regions indicate the acceptances of the CMS/TOTEM detectors relevant for the energy (upper figure) and multiplicity (lower figure) measurements. Note the logarithmic vertical scale.*

# 4 Classifying diffractive interactions

## Physics objects for classification

The physics objects that signify diffraction consist of *leading protons*, *rapidity gaps*, *multiplicity, and transverse energy flows* (see Figs. 5).

To classify proton-proton interactions at the LHC, the following input information is used for the subsequent analyses (Table 2);[18]

- *particle flows* by TOTEM $T1_{R/L}$, $T2_{R/L}$ spectrometers and CMS $FSC_{R/L}$ counters at *±60* to *±140 m* from IP5 [5],
- *transverse energy detection* by the CMS Barrel and End Cap Calorimetry, $HF_{R/L}$, $CASTOR_{R/L}$ and $ZDC_{R/L}$ calorimeters,
- *neutral particle detection* by the CMS $ZDC_{R/L}$ calorimeters.

**Table 2**. *The transverse energy (GeV) and multiplicity (mip)[19] inputs for the event classification algorithms.*

| Variable | Comments |
|---|---|
| E_zdcl | ZDC energy left |
| E_casl | CASTOR energy left |
| E_hfl | HF energy left |
| t2ml | T2 multiplicity left |
| t1ml | T1 multiplicity left |
| fwdm1l | FSC multiplicity left planes 1-2 |
| fwdm2l | FSC multiplicity left planes 3-8 |
| fwdm3l | FSC multiplicity left planes 9-10 |
| fwd1stl | 1st FSC plane hit left |
| fwdmaxl | FSC plane with the maximum amount of hits left |
| E_zdcr | ZDC energy right |
| E_casr | CASTOR energy right |
| E_hfr | HF energy right |
| t2mr | T2 multiplicity right |
| t1mr | T1 multiplicity right |
| fwdm1r | FSC multiplicity right planes 1-2 |
| fwdm2r | FSC multiplicity right planes 3-8 |
| fwdm3r | FSC multiplicity right planes 9-10 |
| fwd1str | 1st FSC plane hit right |
| fwdmaxr | FSC plane with the maximum amount of hits right |
| endc_l | CMS endcap energy left |
| endc_r | CMS endcap energy right |
| barrel | CMS barrel energy |

The CMS and TOTEM detectors will be combined for covering central diffractive and hard single diffractive physics processes consisting of the physics objects listed above. The CMS trigger system will select leptons and jets within the pseudorapidity range of $|\eta| \leq 2.5$, the single and multiple jet triggers within $|\eta| \leq 5$, the missing transverse energy trigger, and very forward jet triggers within $5.3 \leq |\eta| \leq$

---

[18] In present analysis, the leading protons (see Refs. [15-17]) are not used for classifying the diffractive events.
[19] The particle multiplicities are expressed in terms of *mips* = minimum ionizing particles.

*6.6* (CASTOR$_{R/L}$). The energetic neutrons and γ's at zero degree scattering angles, are triggered by the Zero Degree Calorimeter (ZDC$_{R/L}$).[20]

Physics Simulation Tools

The physics data samples for the proton-proton scattering events at *√s = 14 TeV* in:

- single diffraction (SD)
- double diffraction (DD)
- central diffraction (CD)
- non-diffractive processes (ND)

are generated by using the PYTHIA[21] [18], PHOJET [19] and GEANT [20] simulation packages. For central diffraction, a special construction Monte Carlo program based on PHOJET is used[22].

A number of comparisons between the different event generators are available [21] and show that there are very significant differences in both modelling proton-proton interactions in general, and in predicting inclusive observables, such as particle multiplicities and transverse energies at the LHC energies. On the basis of earlier analyses, we have chosen PYTHIA for generating SD and ND events and PHOJET for generating the DD and CD event samples.

The dynamical models used in PYTHIA and PHOJET are very different. While PYTHIA effectively uses perturbative QCD for both low and high $p_T$ processes [18], PHOJET is based on the Dual Parton Model (DPM) [22] with a predefined $p_{Tmin}$ signifying a transition from soft to hard physics. Both PYTHIA and PHOJET describe the experimental data below *√s ≈ 800 GeV* adequately, but in approaching Tevatron energies marked discrepancies emerge (see Table 1 and Ref. [21]). For the SD and ND event categories, PYTHIA predicts about *40%* higher average particle multiplicities in the T1 and T2 regions compared to PHOJET (see Figs. 5).

Datasets

For the following analysis, *12,000* events of each category (SD1, SD2, DD, CD, and ND, see Section 3) were generated using either PYTHIA or PHOJET (Table 3). After Monte Carlo event generation, the data was subjected to the GEANT detector simulation to produce the final dataset. To improve the classification accuracy and to facilitate learning of the data, the SD events were divided into two classes: SD1 and SD2, in which either the beam-1 proton (circulating anti-clockwise in the LHC ring) or the beam-2 proton (circulating clockwise in the LHC ring) is excited to a higher mass state with the proton quantum numbers, i.e. for which the diffractive system ends up on the right or left hand side of the rapidity distributions depicted in Figs. 5 (pages 10 -13). Each dataset was further sub-divided into a training data of *10,000* events and a test data of *2,000* events. The algorithms were trained using the training data and their performance was validated using the test data. The test data is presented to the algorithms only after the training phase is completed and can thus be used to verify the generalization capability of the classifiers.

---

[20] The forward CMS HF calorimeters complement the coverage for forward jets.
[21] The PYTHIA Monte Carlo does not include central diffractive scattering (CD).
[22] The Monte Carlo program can also be used for special studies on low-mass diffractive states: N* → pπ$^o$, nπ$^+$, Δ$^{++}$π$^-$. A detailed description of the program can be found in [23].

**Table 3**. *Summary of the Monte Carlo simulators used to generate the datasets.*

| Event type | Generator |
|---|---|
| **SD** | PYTHIA 6.205 |
| **DD** | PHOJET 1.12 |
| **CD** | PHOJET 1.12 |
| **ND** | PYTHIA 6.205 |

# 5   Multivariate techniques

Multivariate algorithms such as neural networks (NN), Fisher discriminants (FD), kernel estimation methods (KM), or support vector machines (SVM) have been introduced in order to improve the physics analysis in a multidimensional parameter space. More recently, gene expression programming (GEP) [24] has been used also in high energy physics analysis [25]. This novel approach is of special interest in the subsequent feasibility study. The traditional Boolean method is based on successive event selections (cuts) that tend to be on *ad hoc* basis and suppress both background *and* the signal.

In this paper, a multivariate analysis of the proton-proton collisions at the LHC energies is carried out by using gene expression programming [24, 29], neural networks [27] and support vector machine [28] approaches with large number of inputs available for classifying the different categories: SD, DD, CD and ND[23]. In order to assess the systematic uncertainties in event selection, the classification results obtained by the three methods are compared in detail.

Gene Expression Programming

Gene expression programming (GEP) [24,29] is a recently introduced evolutionary algorithm which is used to evolve expression trees such as the one depicted in Figure 6. GEP is distinguished from other evolutionary algorithms by having separate representations and structures for the genotype (the chromosome) and the phenotype (the expression tree) of an individual. That is, the algorithm is used to evolve simple, easy to manipulate, linear chromosomes represented as text strings, which in turn encode the more complex expression trees of various shapes and sizes. GEP can be seen as a combination of genetic algorithms and genetic programming; the former evolves binary strings while the latter is used to optimize tree-like entities.

---

[23] For the present analysis *23* input variables are used (Table 2).

The nodes of a GEP expression tree can either be members of a function set or a terminal set. The former consists of all the mathematical functions the algorithm is allowed to use, while the latter contains input variables and random constants. The chromosome of GEP is expressed in the so called *Karva* language. Each chromosome consists of a head of length $h$ and a tail of length $h(n-1)+1$, where $n$ is the number of arguments for the function with the largest number of arguments in the function set. The head can consist of elements of both the function set and the terminal set, while the tail can only contain elements of the terminal set. This structure ensures that every *Karva* string is a valid expression tree.

A chromosome could, for example, be the following:

$$-*/aQ\mathbf{bcaacb,} \tag{7}$$

where the tail is bold faced, lower case letters are members of the terminal set, and Q denotes a unitary function. This would be translated into the expression tree of Figure 6 by reading the *Karva* string from left to right. Note, that in this particular case, some of the tail section is not used in the final expression tree at all.

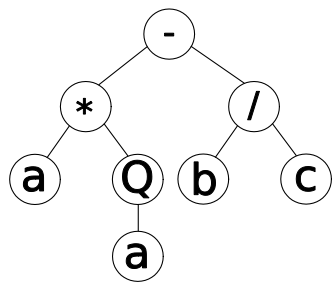

**Figure 6** *Karva expression (Eq. 7) translated into an expression tree.*

The goal of the GEP algorithm is to evolve the expression tree, encoded by the *Karva* expression, so that the output of the tree maximizes a predefined performance criterion, the fitness function. The expression trees evolved by GEP are here used as classifiers, i.e. the objective is to maximize the classification accuracy of the output of the tree given a set of input vectors. In this paper, each tree is used for binary classification (*signal/background* separation) only, by introducing a threshold (= *0.5*) for the output of the tree. When the output is equal to or greater than the threshold, the input vector is classified as *signal* and below the threshold as *background*.

The sensitivity/specificity fitness function is employed in this analysis, i.e.

$$SS = SE \cdot SP. \tag{8}$$

The sensitivity is then given as

$$SE = \frac{TP}{TP+FN}, \tag{9}$$

where *TP* is the number of true *positive* classifications and *FN* is the number of false *negative* classifications by a certain expression tree. It should be noted, that in particle physics community, sensitivity is usually referred to as *efficiency*.

The specificity is given as

$$SP = \frac{TN}{TN + FP}, \qquad (10)$$

where *TN* is the number of true *negative* classifications, and *FP* the number of false *positive* classifications. With this fitness function, an individual is penalized for both false negatives and false positives.

The GEP algorithm attempts to maximize the fitness function (Eq. 8) by evolving a population of chromosomes using a set of genetic operators. These include, for example, mutation where an element of a chromosome is randomly replaced by another element, and recombination where two chromosomes produce an offspring by exchanging a sequence of genetic material. The complete GEP algorithm includes a dozen different genetic operators and a number of small improvements to the basic concept, such as multigenic chromosomes and random numerical constants.[24]

The GEP classifiers are here evolved using GeneXproTools 4.0 by Gepsoft. We used the default function set of this implementation which consists of *30* functions: different basic arithmetic operations, mathematical functions, logic operators and comparison functions. The terminal set consisted of *23* input variables (see Table 2, page 14) and *8* random numerical constants. Multigenic chromosomes with four genes and head size of *h = 10* were used. The population of chromosomes had *30* individuals. By fine-tuning of the parameters, no significant improvement in the performance of the algorithm was achieved. Data normalization was also studied, but no meaningful improvements on the performance of the GEP algorithm were obtained. The subsequent analysis is, therefore, based on data that is *not* normalized.

An advantage of the GEP approach is due to its transparency in comparison with other types of multivariate analyses, such as neural networks, for example. By studying the GEP expression trees, the variables that are important for the event classification can be mapped out. This, in turn, allows the key detectors to be identified. Furthermore, if the function set is chosen to be simple enough, important physical interpretations can be assigned for the expression trees.

Neural Networks

Neural networks (NN) are adaptive data modelling tools inspired by the functional model of the human brain [27]. They can be used in a wide array of tasks such as classification, time-series analysis and decision making. Since about a decade, the neural networks are widely used in high-energy physics data analysis. In this paper, a particular type of feed-forward neural networks, called the multi-layer perceptron (MLP) network is used. The MLP network consists of an input layer, output layer and one or more hidden layers of neurons. In this work, only the MLP networks, with a single hidden layer, are considered.[25] When information propagates through the network in the forward direction, the weighted sum of the activation levels of the input neurons is fed into a hidden layer of $N_{hid}$ neurons. The activation level of the hidden neurons is determined by a transfer function $f$ whose output is, in turn, fed into the output layer. The process can be described by an equation

$$\mathbf{y} = \mathbf{B}f(\mathbf{A}\mathbf{x} + \mathbf{a}) + \mathbf{b} \qquad (11)$$

where **y** and **x** are the output and input vectors respectively, while **A** and **B** are weight matrices. This formulation also includes the bias vectors **a** and **b**, and the nonlinear transfer function $f$ is taken component-by-component on the input for the hidden layer $\mathbf{A}\mathbf{x} + \mathbf{a}$. The transfer function, $f(\mathbf{x}) = \tanh(\mathbf{x})$, is used in this analysis. The MLP-network architecture is illustrated in Figure 7.

---

[24] For a detailed description of the algorithm, see Ref. [29].

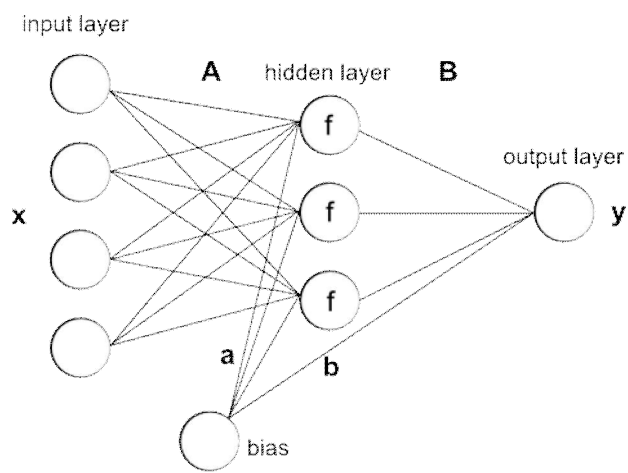

**Figure 7** *Schematic representation of the MLP network (Eq. 11).*

An MLP network is trained by the so called back-propagation[26] algorithm [31] which minimizes the squared error between the training data and the output of the network

$$E = \frac{1}{2}\sum_i (y_i - t_i)^2, \qquad (12)$$

where $t_i$ is the $i^{th}$ component of the training data vector **t**. The error is minimized by using an optimization method based on gradient descent in the space of network weights and biases. The NNs used in this paper are trained using a batch training procedure where all training instances are first presented to the network, and only after this, the sum of individual updates are used to update the network weights. Since the result of the training process depends on the random initialization of the network weights, the learning procedure is repeated ten times and the classifier that leads to the highest classification accuracy is used in further analysis.

---

[25] These have been shown to be universal function approximators, provided that there is a sufficient amount of neurons in the hidden layer [30].
[26] The name of the algorithm originates from the fact that in this particular case the chain rule of differentiation can be regarded as an error propagating backwards in the network.

The Matlab R2009a Neural Network Toolbox is used to train the NNs. The training algorithm is the Levenberg-Marquardt back-propagation, which is a powerful combination of the gradient descent and Gauss-Newton algorithms. The number of hidden neurons is set to $N_{hid}=18$. An early stopping strategy is employed, with *20%* of the training dataset set aside as a validation dataset. The performance of the network on the validation set is monitored; when the performance starts to decrease, this is regarded as a signal of overfitting, and the training is stopped.

It is well known, that the performance of NNs can be greatly improved by normalizing the data [32]. In this study, each component of the NN input vectors is first normalized by using logarithmic normalization: $x \mapsto \log(1+x)$. A linear scaling to the interval [*-1,1*] follows as

$$x \mapsto 2\frac{x - x_{min}}{x_{max} - x_{min}} - 1 \;, \tag{13}$$

where $x_{min}$ and $x_{max}$ are the minimum and maximum values of *x* in the training data. This transformation is, naturally, also applied to the testing data.

## Support Vector Machines

Support Vector Machines (SVMs) [28] have become a popular multivariate analysis tool in high energy physics. SVMs are mostly used for classification tasks, but they can also be applied to e.g. regression. The main idea in SVMs is to find a hyperplane that separates two different data samples, representing different classes, with the largest possible margin. The margin is defined as the distance from the hyperplane to the closest data points. It is not always possible to find such a plane and, therefore, the data points are usually projected nonlinearly into a higher dimensional space before finding the optimal hyperplane.

The standard SVM algorithm solves a binary classification problem, where the goal is to find a function $f : \mathbf{R}^n \to \pm 1$ that correctly classifies the data points $\mathbf{x}_i (i = 1...N)$ into classes $y_i = 1$ and $y_i = -1$. A hyperplane can be written down as $(\mathbf{w} \cdot \mathbf{x}) + b = 0$, where $\mathbf{w}$ is called the *weight vector*. The corresponding decision function for the hyperplane becomes

$$f(\mathbf{x}) = \text{sign}[(\mathbf{w} \cdot \mathbf{x}) + b]. \tag{14}$$

The goal is to maximize the margin which is achieved by minimizing $\frac{1}{2}\|\mathbf{w}\|^2$ [28]. From the decision function $f(\mathbf{x})$, the following constraint is obtained for the optimization task

$$y_i[(\mathbf{w} \cdot \mathbf{x}_i) + b] > 0, \forall i \,. \tag{15}$$

Furthermore, a canonical hyperplane is defined, for which $(\mathbf{w} \cdot \mathbf{x}) + b = 1$, for the closest data point on one side of the plane and for which $(\mathbf{w} \cdot \mathbf{x}) + b = -1$, for the closest point on the other side. With this canonical hyperplane, the constraint can be written as

$$y_i[(\mathbf{w} \cdot \mathbf{x}_i) + b] \geq 1, \forall i. \tag{16}$$

In order to make the data separable by hyperplanes, a mapping $\Phi$ to a higher dimensional space can be conducted. However, the exact mapping function is not required, but it is sufficient to find a kernel function, K, that defines the inner product in the transformed space:

$$K(\mathbf{x}_i, \mathbf{x}_j) = \Phi(\mathbf{x}_i) \cdot \Phi(\mathbf{x}_j). \tag{17}$$

The kernel function is usually chosen as one of the following three functions

- Polynomial kernel: $K(\mathbf{x}_i, \mathbf{x}_j) = (\mathbf{x}_i \cdot \mathbf{x}_j + a)^d$
- Gaussian Radial Basis Function kernel: $K(\mathbf{x}_i, \mathbf{x}_j) = \exp(-\gamma \|\mathbf{x}_i - \mathbf{x}_j\|^2)$
- Sigmoid kernel: $K(\mathbf{x}_i, \mathbf{x}_j) = \tanh(k\mathbf{x}_i \cdot \mathbf{x}_j + \theta)$,

with constants $a$, $d$, $\gamma$, $k$ and $\theta$.

Since real data is challenged by backgrounds, a complicated kernel function may be required leading into problems with over-fitting. This can be avoided by introducing a slack variable $\xi_i$ in the constraint function:

$$y_i[(\mathbf{w} \cdot \mathbf{x}_i) + b] \geq 1 - \xi_i, \forall i. \tag{18}$$

The SVM implementation called BSVM [33] is used in this analysis. As a kernel, the Gaussian Radial Basis Function (the default kernel in BSVM) is adopted. Parameter $\gamma$ is optimized with a tool that is included in the BSVM. The performance of the SVM can be significantly improved by linearly normalizing the data. A logarithmic normalization before the linear normalization[27] was found to further improve the results. This could be due to a relatively few variables receiving very large values compared to the mean of the variables. A plain linear normalization allows these abnormally high values to dominate the classification since all other values become close to zero. Therefore, a logarithmic normalization $x \mapsto \log(1 + x)$ is first carried out, then the values are scaled by dividing them with the maximum reached by each component $x \mapsto x / x_{max}$.

## Multi-class classification

The classification algorithms are usually designed for binary problems, where the goal is to distinguish two classes [34,35]. Therefore, multi-class classification tasks are often reduced into several binary problems. Several different techniques for the reduction exist; here three of them: *one-against-all*, *one-against-one* and *ordered binarization* are introduced.

---

[27] A linear normalization tool serves as a default solution in the BSVM implementation package.

*One-against-all*

The one-against-all method reduces a *c*-class problem into *c* binary problems. The idea is to train, for each class, a classifier separating the class from the rest of the classes. This method is used here with the GEP algorithm. The *c* different classifiers in GEP are trained to output a value equal to or greater than the threshold value of *0.5* for the events that are being separated from rest of the data (label *1*). For the rest of the events (label *0*), the classifiers should give values less than *0.5*. In case exactly one classifier outputs label *1*, it is obvious that the event should be labelled according to this classifier. However, if more than one or no classifiers output label *1*, it is not clear how the event should be labelled. In this case, the label is simply chosen according to the classifier that outputs the largest value.

*One-against-one*

The one-against-one method forms a classifier for each *2*-class combination resulting in *c(c-1)/2* different classifiers. An event is run through all the classifiers, and each classifier gives one point to either one of the two classes. The class with the highest score is chosen as the output class. This system is also known as the round robin system. The one-against-one method is used in SVM since it is one of the default multi-class classification options in the BSVM, which is the SVM implementation used in the present analysis. If any pair of classes gets the same score, BSVM chooses the one with a smaller index [35].

*Ordered binarization*

In ordered binarization, classifiers similar to the one-against-all classifiers are formed, except that the classes are dropped out one by one. Assuming the classes *A, B, C* and *D*, the following classifiers would be formed: *A vs. {B, C, D}*, *B vs. {C, D}* and *C vs. D*. In the one-against-all method, the order in which the classifiers are applied is irrelevant. On the contrary, here the classification is always started from the "*A vs. the rest*" classifier. The classifier that first outputs label *1* determines the class for the event. The motivation for this approach is that, e.g. with our dataset the algorithms are able to distinguish ND events with a very high efficiency and purity. Therefore, the ND events are first separated from the rest. This approach simplifies the remaining classification task, and also avoids the problem of draw situations present in the previous two methods. On the other hand, if one of the first classifiers in ordered binarization falsely labelled an event as *1*, the rest of the classifiers cannot have their opinions on the event anymore. Ordered binarization is used with all the three different algorithms (SVM, NN and GEP) and is performed starting with ND followed by CD, SD1, SD2 and DD.

In addition to these three methods, a further special method based on neural networks is investigated. In that method, only one network with *c* output nodes corresponding to the *c* different classes is required. The network is trained to give zero outputs for all the false classes, and output exactly one for the correct class. An event to be classified is fed through this network, and the class is determined according to the output node with the highest value.

## Self-organizing maps

The self-organizing map (SOM) is a computational method which can be used, e.g. for dimensionality reduction and data visualization [36]. With a SOM, a non-linear mapping of the analysed *23*-dimensional space to a two dimensional map is achieved.

The map consists of *n* by *m* nodes which all contain a model vector. The nodes are usually arranged on a hexagonal grid. The mapping of an input vector is conducted by going through all nodes and calculating Euclidean distances between model vectors and the input vector. The node with the smallest distance to the input vector is called the best matching unit (BMU) and the input vector is mapped to this node.

The training process is conducted similarly by calculating BMUs. The model vectors are first initialized randomly[28]. After the initialization, a set of input vectors is given to the algorithm. The algorithm calculates the BMU for each input vector, **x**, and adjusts the model vector, **m**$_c$, in the BMU to make it more similar to the input vector. The algorithm also adjusts the neighbouring model vectors according to some neighbourhood function $h_{ci}$ where *i* denotes the index of the model vector to be adjusted and *c* the index of the BMU model vector. The update rule for the model vectors is

$$\mathbf{m}_i(t+1) = \mathbf{m}_i(t) + \alpha(t) h_{ci} (\mathbf{x} - \mathbf{m}_i(t)) \qquad (19)$$

where *t* is the training iteration and $\alpha(t)$ is the learning rate. Usually a Gaussian neighbourhood function is chosen and thus a model vector is adjusted the more the closer it is to the BMU on the grid. If the neighbourhood size is set to zero (only the BMU is adjusted), the SOM algorithm reduces into the *k*-means algorithm. The SOM algorithm is unsupervised, i.e. it does not require a set of vectors with predefined projections on the map; the map can be trained with the same vectors that are to be mapped. SOM tends to preserve the topological properties of the original space, i.e. the vectors that are close to each other in the original space tend to remain close neighbours in the resulting map as well. This feature of the SOM algorithm allows it to form clusters of similar events on the map.

*Analysis of the pp-events with the self-organizing map*

The SOM algorithm is here employed to visualize how the different event categories are separated in the multivariate analysis. A SOM is trained with *60,000* PYTHIA or PHOJET simulated events (*12,000* of each type) by using the SOM Toolbox 2.0 [37]. The default parameters of this implementation were used except for data normalization for which a logarithmic normalization was found to give the best results. The different event categories are mapped on the SOM (Figure 8), with colour codes to identify the event categories: red for the SD1, green for the SD2, blue for the ND, black for the DD and yellow for the CD events. The larger the colour patch on a node the more events are mapped to the node. The map clearly demonstrates that the non-diffractive events are easily identified; they are basically all clustered at the bottom of the map. Similarly, the CD events are rather well separated from the other diffractive event categories. The most significant overlap occurs between the SD and DD events.

---
[28] More sophisticated initialization methods also exist.

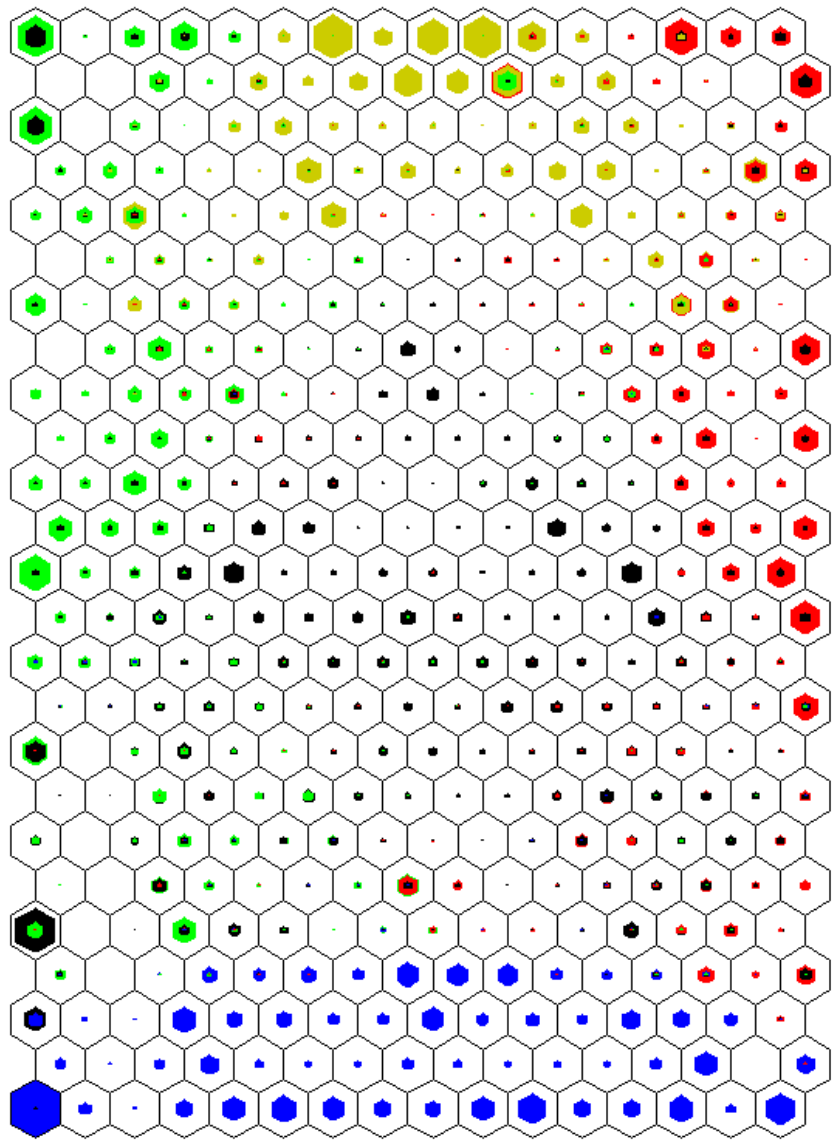

**Figure 8**. *The Self Organizing Map (SOM) of the pp event categories. The following colours are used to label different event categories: red = SD1, green = SD2, blue = ND, black = DD and yellow = CD.*

The SOM technique can also be used as an efficient visualization tool to differentiate between the characteristics of the different map nodes in the original *23*-dimensional space (for the *23* input variables, see Table 2, p. 14). In Figure 9, each input dimension is shown in a separate component plane. Red colour on a node of a component plane denotes that the variable usually receives large values among the events mapped to that node. For example, the plane in the bottom right corner, representing the CMS barrel calorimeter values, indicates that the CMS barrel calorimeter usually detects large energy depositions from the events that are mapped to the bottom and to the top-centre of the map. These regions mainly contain ND events and CD events. That is, the ND and CD events tend to release large amounts of energy within the CMS barrel.

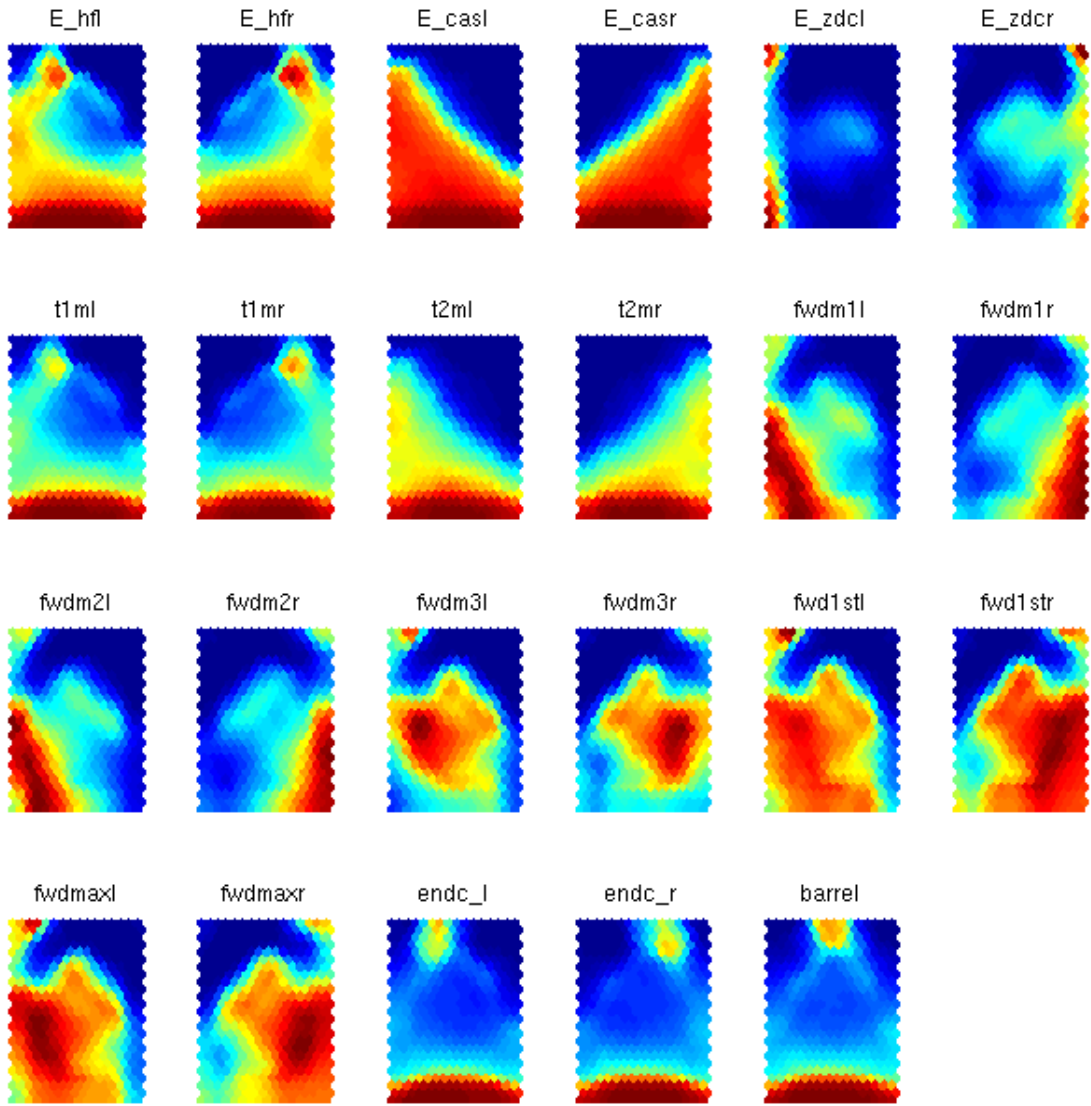

**Figure 9** *The component planes that show the SOM nodes in the original 23-dimensional space used for the analysis. The colours indicate the magnitudes of the variables.*

By examining the component planes of each dimension, the event characteristics that will most efficiently distinguish different types of events can be chosen. For example, the FSC planes 9-10 (fwdm3l/fwdm3r) are very useful for identifying DD events, since these variables have high values in the central part of the map where the DD events are located. Event characteristics can be also examined with the simulated energy and multiplicity flow distributions shown in Figures 5 (pp. 10 -13) and it is instructive to compare these different event qualities. The component planes can also be used for deducing correlations between different variables. For example, the 1st FSC plane hit (fwd1stl/fwd1str) and the FSC plane with the maximum amount of hits (fwdmaxl/fwdmaxr), seem to provide almost the same information. One may use this sort of inference for eliminating redundant variables in order to ease the task of a learning algorithm.

An analogy between the SOM technique utilized above and a histogram analysis can be drawn, since the nodes of a SOM correspond to the bins of a histogram. However, the values that a bin represents are not read from the axis around the bin, but from the component planes. The SOM algorithm becomes very useful in case several types of events are analyzed simultaneously. Plotting these events into the same histograms, easily leads into ambiguities. With the SOM algorithm, this problem is solved by assigning the events to the nodes of the map so that similar events get assigned to the same node, and by constructing a component plane for each dimension. These planes represent the typical values (determined by the model vectors) that the dimensions exhibit in different nodes.

# 6 Event classification results

Tables 4, 5 and 6 summarize the efficiencies and purities of the three event classification methods under study. The values are calculated using the test data consisting of a sample of *2,000* events of each type. To simplify the presentation, we have combined the results of the two SD classes as a single class. The results are calculated so that the algorithms are not penalized for mixing the SD1 and SD2 classes. In successive rows (Table 4), the probability of different methods (GEP, SVM, NN) to identify the 'correct' process, and the 'false' ones are shown. For example, in row one of the GEP table (a), the first number indicates that the gene expression programming allows *83.35%* of the double diffractive (DD) proton-proton scattering events to be correctly identified among the inclusive (SD+DD+CD+ND) proton-proton scattering events simulated by PYTHIA and PHOJET. In successive columns of row one, it is shown that the method incorrectly assigns *16.25%* of the DD events as SD, *0.40%* as CD and *0.00%* as ND type.

The average efficiencies of the different methods calculated as the average of the diagonal elements of the matrices in Table 4 are summarized in Table 5. These values represent the probability of correctly classifying an event belonging to a randomly selected class.

The purities of the different classes with the various methods are shown in Table 6. These values represent the probability that an event classified to some class really belongs to that particular class. For example, it is seen that *96.72%* of the events classified as DD, using GEP with ordered binarization, are real DD events. In Table 4(a) it is seen that the deviation from a perfect result is mostly caused by some SD events being classified as DD events.

**Table 4.** *Performance of gene expression programming (GEP), support vector machine (SVM) and neural network (NN) based multivariate analyses in classifying the proton-proton event categories into single diffractive (SD), double diffractive (DD), central diffractive (CD) and non-diffractive (ND) event categories. Each row of the tables represent the correct event class and each column the class identified by the classifier.*

(a) Gene Expression Programming (GEP): "ordered binarization"

| Class | DD | SD | CD | ND |
|---|---|---|---|---|
| **DD** | 83.35 | 16.25 | 0.40 | 0.00 |
| **SD** | 2.58 | 89.75 | 6.80 | 0.88 |
| **CD** | 0.00 | 1.15 | 97.50 | 1.35 |
| **ND** | 0.25 | 0.40 | 0.00 | 99.35 |

(b) Gene Expression Programming (GEP): "one-against-all"

| Class | DD | SD | CD | ND |
|---|---|---|---|---|
| **DD** | 87.90 | 10.20 | 0.90 | 1.00 |
| **SD** | 9.18 | 82.22 | 8.25 | 0.35 |
| **CD** | 0.00 | 6.10 | 92.60 | 1.30 |
| **ND** | 7.75 | 0.80 | 0.00 | 91.45 |

(c) Support Vector Machine (SVM): "ordered binarization"

| Class | DD | SD | CD | ND |
|---|---|---|---|---|
| **DD** | 86.80 | 12.75 | 0.45 | 0.00 |
| **SD** | 1.90 | 94.92 | 3.15 | 0.03 |
| **CD** | 0.00 | 4.55 | 95.45 | 0.00 |
| **ND** | 0.10 | 0.25 | 0.00 | 99.65 |

(d) Support Vector Machine (SVM): "one-against-one"

| Class | DD | SD | CD | ND |
|---|---|---|---|---|
| **DD** | 86.65 | 12.95 | 0.35 | 0.05 |
| **SD** | 1.98 | 94.95 | 3.02 | 0.05 |
| **CD** | 0.00 | 3.70 | 96.30 | 0.00 |
| **ND** | 0.15 | 0.25 | 0.00 | 99.60 |

(e) Neural Networks (NN): "ordered binarization"

| Class | DD | SD | CD | ND |
|---|---|---|---|---|
| **DD** | 87.60 | 12.05 | 0.35 | 0.00 |
| **SD** | 2.15 | 95.20 | 2.58 | 0.07 |
| **CD** | 0.00 | 4.25 | 95.75 | 0.00 |
| **ND** | 0.15 | 0.25 | 0.00 | 99.60 |

(f) Neural Networks (NN): "5 outputs"

| Class | DD    | SD    | CD    | ND    |
|-------|-------|-------|-------|-------|
| DD    | 86.90 | 12.65 | 0.45  | 0.00  |
| SD    | 1.90  | 95.15 | 2.88  | 0.07  |
| CD    | 0.00  | 3.80  | 96.20 | 0.00  |
| ND    | 0.15  | 0.40  | 0.00  | 99.45 |

**Table 5.** *The average efficiencies of gene expression programming (GEP), support vector machine (SVM) and neural network (NN) based multivariate analyses in classifying the proton-proton event categories.*

| Method                  | <Efficiency> |
|-------------------------|--------------|
| GEP ordered binarization | 92.49        |
| GEP one-against-all     | 88.54        |
| SVM ordered binarization | 94.21        |
| SVM one-against-one     | 94.38        |
| NN ordered binarization | 94.54        |
| NN 5 outputs            | 94.42        |

**Table 6.** *Purities of gene expression programming (GEP), support vector machine (SVM) and neural network (NN) based multivariate analyses in classifying the proton-proton event categories into single diffractive (SD), double diffractive (DD), central diffractive (CD) and non-diffractive (ND) event categories.*

| Method                  | DD    | SD    | CD    | ND    |
|-------------------------|-------|-------|-------|-------|
| GEP ordered binarization | 96.72 | 83.45 | 93.12 | 97.81 |
| GEP one-against-all     | 83.85 | 82.78 | 91.01 | 97.18 |
| SVM ordered binarization | 97.75 | 84.40 | 96.37 | 99.97 |
| SVM one-against-one     | 97.61 | 84.89 | 96.61 | 99.90 |
| NN ordered binarization | 97.44 | 85.19 | 97.04 | 99.92 |
| NN 5 outputs            | 97.70 | 84.96 | 96.66 | 99.92 |

By studying the "*ND vs. the rest*" GEP classifier (see p. 18), the relative importance of different CMS/TOTEM detectors can be assessed in discriminating the non-diffractive events. It is noted, for example, that the variables related to the CASTOR calorimeter and to FSC planes 9 to 10, are not used by the algorithm, i.e. they do not seem to have additional discrimination power vs. ND events. It is instructive to compare this observation with the SOM component planes in Fig. 9.

Out of the methods analyzed, NN with ordered binarization achieves the best efficiency. It also achieves the highest purity for the SD and CD classes. For the DD and ND classes, the best purity is given by SVM with ordered binarization. Nevertheless, the performance of all the algorithms, with the notable exception of GEP with one-against-all multi-classification, seem to achieve very similar results which is an indication that the remaining mixing between the classes is due to specific characteristics of the data, and not a shortcoming of the algorithms or the training process. This hypothesis is also supported by an analysis of the misclassified events of the different algorithms which reveals that a large majority of these events are the same. For example, with the ordered binarization NN 71%-89% of the misclassified events are shared with the five other algorithms. By plotting these misclassified events to a SOM map, it is seen that they concentrate into the map regions where different event categories overlap, as one might intuitively expect.

All three multivariate analysis algorithms are seen to identify the diffractive event categories from the non-diffractive one with high efficiency *(ε ≥ 99%)*. The CD events are also easily separated among the diffractive event categories *(ε ≥ 95%)*. Separation of the SD and DD event categories is, however, more challenging and the efficiencies of the DD class calculated for different methods vary between *ε ≈ 83 – 88%* as a result of many DD events being incorrectly classified as SD events. Consequently, the purity of the SD class is around *84%* with all the methods while other classes achieved purities in the order of *96 – 99%*. These results are in good accordance with the SOM analysis presented in Section 5.

In case some of the foreseen forward detector systems fail to be in use at the LHC start-up, the classification efficiencies within the diffractive event classes decrease significantly. As an example, by removing the T1 detectors, the CASTOR calorimeter on the right, and the FSC detectors, and analyzing the remaining data with the ordered binarization NN, the DD event discrimination drops below *70%*, SD efficiency to about *88%* and the CD efficiency to about *81%*. The ND events are still identified with the relatively high efficiency of *99.4%*. This is mainly thanks to T2 and the CMS calorimeters as revealed by the SOM analysis in Section 5. Furthermore, one can deduce from the SOM component planes in Figure 9 that the decrease in the efficiency of the DD classification is due to the removal of the FSC planes 9-10.

The effect of removing different sets of detectors from the algorithms was also investigated. In case only the base line CMS detectors are used (i.e. the T1, T2 and FSC detectors are removed from the analysis), the average efficiency of the ordered binarization NN was found to decrease to about 85%. If only T1 and T2 telescopes were removed, an average efficiency of 94.0% was obtained. This underlines the importance of the proposed FSCs for an efficient event classification. Furthermore, if only particle multiplicity data from the T1 and T2 spectrometers were used, the average efficiency plunged down to about 64%; by adding the FSCs, the number increased to ~ 86%.

# 7 Conclusions

A set of novel multivariate analysis based event classification methods for detecting diffractive interactions at the LHC has been developed. Diffractive (SD, DD, CD) and non-diffractive (ND) event samples were first generated by the PYTHIA and PHOJET Monte Carlo event generators, then passed through GEANT based detector simulation program. The non-diffractive background is easily rejected by the all three multivariate techniques investigated in this analysis, and an efficient classification of the single diffractive (SD), double diffractive (DD) and central diffractive (CD) event categories is achieved.

The analysis clearly demonstrates the key importance of the FSCs in classifying the diffractive events. Moreover, when the CMS detectors are excluded from the process, the average classification efficiencies drop dramatically.

It should be noted, that the event classification results introduced in this paper depend on the particular Monte Carlo models used to train and test the algorithms. Nevertheless, as long as the models correctly reflect the kinematical constraints (energy-momentum conservation) and the cross features of different event categories, the results should reliably reflect the efficiencies and purities of a real event analysis at the LHC.

Further development is being carried out by the authors to (1) combine the best features of the three techniques into a unified approach for event-by-event analysis, and (2) to develop an unsupervised probabilistic scenario that is less dependent on a particular Monte Carlo model in use. The aim of the ongoing work is to develop an algorithm that can be used to evaluate relative rates of different diffractive event categories and, finally, to optimize the analysis of central diffractive production of $J^{PC} = 0^{++}$ states, such as heavy quarkonia, glueballs, Higgs boson, etc.

# Acknowledgements

Valuable advice from Albert De Roeck, Paul Hoyer, Valery Khoze, Mikhael Ryskin and financial support by the Academy of Finland are gratefully acknowledged.